\documentclass[twocolumn,showpacs,preprintnumbers,amsmath,amssymb]{revtex4}

\usepackage{slashed}

\def\ket#1{|#1\rangle}
\def\bra#1{\langle#1|}

\def\fancys{\mathbb{S}}

\newcommand{\be}{\begin{equation}}
\newcommand{\ee}{\end{equation}}
\newcommand{\bea}{\begin{eqnarray}}
\newcommand{\eea}{\end{eqnarray}}
\newcommand{\ba}{\begin{array}}
\newcommand{\ea}{\end{array}}

\def\bbox{{\,\lower0.9pt\vbox{\hrule \hbox{\vrule height 0.2 cm
\hskip 0.2 cm \vrule height 0.2 cm}\hrule}\,}}
\newcommand{\dsl}{\pa \kern-0.5em /}

\usepackage{graphicx}
\usepackage{dcolumn}
\usepackage{bm}

\begin{document}

\preprint{MIT-CTP-4277}

\title{Unification of Type II Strings and T-duality}

\author{Olaf Hohm, Seung Ki Kwak, and Barton Zwiebach}
\affiliation{%
Center for Theoretical Physics, Massachusetts Institute of Technology,\\
Cambridge, MA 02139, USA.
}%

\begin{abstract}

We present a unified description of the low-energy limits of type II string theories. 
This is achieved by a formulation that doubles the space-time 
coordinates in order to realize the T-duality group $O(10,10)$ geometrically. 
The Ramond-Ramond fields are described by a spinor of $O(10,10)$, which couples to the  gravitational fields via the Spin$(10,10)$ representative of the so-called generalized metric.  
This theory, which is supplemented by a  T-duality  
covariant self-duality constraint, unifies the 
type II theories in that each of them is obtained for a 
particular subspace of the 
doubled space.

\end{abstract}

\maketitle

Superstring theory in ten  
dimensions is arguably the most promising candidate for
a unified quantum mechanical 
description of gravity and other interactions.  
This theory, however, takes different guises. 
For instance, there are two different string theories with maximal supersymmetry, the 
type IIA and the type IIB theory. 
The ten-dimensional superstring theories, together with 11-dimensional supergravity, are different limits of a single underlying theory 
and  are related through a web of dualities (see, e.g., \cite{Schwarz:1996bh}).
The simplest of these dualities is T-duality  
that, for instance, relates type IIA string theory on the circular 
background $\mathbb{R}^{8,1}\times S^{1}$ of radius $R$ to type IIB string theory 
on the same background, but with radius $1/R$.  
 
In its low-energy limit string theory is described by Einstein's theory of general relativity, 
coupled to particular matter fields.  In this description, T-duality 
results  in 
the appearance of the hidden symmetry group $O(d,d)$ upon dimensional 
reduction on a torus $T^{d}$. Moreover, 
the low-energy
limits of type IIA and type IIB give rise to the same theory, 
consistent with their  
equivalence under T-duality~\cite{Bergshoeff:1995as}. 

The general coordinate invariance of Einstein gravity naturally 
explains the presence of the $GL(d)$ subgroup, but the emergence of the 
full $O(d,d)$ upon dimensional reduction requires the precise 
matter couplings predicted by string theory, hinting at a novel geometrical structure.
Recently, a `double field theory' (DFT) has been found which realizes 
a T-duality group prior to dimensional reduction 
\cite{Hull:2009mi, Hohm:2010jy} (see also \cite{Siegel:1993th,Tseytlin:1990nb}).  By doubling the 
space-time coordinates, the low-energy effective action of bosonic string theory 
or, equivalently, of the Neveu-Schwarz--Neveu-Schwarz (NS-NS) sector 
of superstring theory, can be extended to an action that has $O(D,D)$ as a global 
symmetry, where $D$ is the space-time dimension. 

In this Letter we introduce the 
extension to the Ramond-Ramond (RR) sector of type II strings, 
which will lead to a theory that contains all type II theories simultaneously 
in different T-duality `frames'. Here we will not present explicit derivations,  
but a more detailed exposition will  appear elsewhere \cite{ToAppear}. 
Related work has appeared in \cite{West:2010ev,Thompson:2011uw}.

We start by reviewing the NS-NS subsector. It consists of the metric $g_{ij}$, the 
Kalb-Ramond 2-form $b_{ij}$ and the scalar dilaton $\phi$, where $i,j,\ldots=1,\ldots, D$
are space-time indices. The DFT is formulated in terms of a dilaton density $d$, which 
is related to $\phi$ via the field redefinition $e^{-2d}=\sqrt{g}e^{-2\phi}$, $g=|\det{g}|$, 
and the `generalized metric'  
 \be\label{firstH}
  {\cal H}_{MN} =  \begin{pmatrix}    g^{ij} & -g^{ik}b_{kj}\\[0.5ex]
  b_{ik}g^{kj} & g_{ij}-b_{ik}g^{kl}b_{lj}\end{pmatrix} \;,
 \ee
which combines $g$ and $b$ into an $O(D,D)$ covariant tensor with indices 
$M,N,\ldots=1,\ldots,2D$. All fields depend on the doubled coordinates 
$X^{M}=(\tilde{x}_{i},x^{i})$. We can regard ${\cal H}$ as the fundamental field,
taking values in $SO(D,D)$ and satisfying ${\cal H}^{T}={\cal H}$, and view  
(\ref{firstH}) as just a particular parametrization. 
The action can be written as   
 \be\label{actR}
   S =   \int dx \, d\tilde x  \, e^{-2d} \, {\cal R}({\cal H},d)\;,
 \ee
where ${\cal R}({\cal H},d)$ is an $O(D,D)$ invariant scalar, cf.~(4.24) in
the second reference of \cite{Hohm:2010jy}, and we use the 
short-hand notation $dx=d^{D}x$. The action is invariant under the 
gauge transformations 
 \bea\label{manifestH}
 \begin{split}
  \delta_{\xi}{\cal H}_{MN} &=  \xi^{P}\partial_{P}{\cal H}_{MN}
  +2\big(\partial_{(M}\xi^{P} -\partial^{P}\xi_{(M}\big)\,{\cal H}_{N)P} \;, \\
  \delta_{\xi} d &=  \xi^M \partial_M d - {1\over 2}  \partial_M \xi^M \,,
 \end{split}
 \eea
with the derivatives $\partial_{M}=(\tilde{\partial}^{i},\partial_{i})$. Here,  
$O(D,D)$ indices $M,N$ are raised and lowered with the invariant metric 
 \be\label{ODDmetric}
  \eta_{MN}  =  \begin{pmatrix}
    0&1 \\1&0 \end{pmatrix}\,,
 \ee
and (anti-)symmetrizations are accompanied by the combinatorial factor $\tfrac{1}{2}$. 
The consistency of the above theory 
requires the constraint 
 \be\label{ODDconstr}
   \partial^{M}\partial_{M}A = \eta^{MN}\partial_{M}\partial_{N}A = 0\,, \quad
   \partial^{M}A\,\partial_{M}B = 0\;, 
  \ee
for all fields and parameters $A$ and $B$. This constraint implies that locally
the fields depend only on half of the coordinates, and one 
can always find an $O(D,D)$ transformation into a frame in which the fields depend 
only on the $x^{i}$. If one drops the dependence on the `dual coordinates' $\tilde{x}_{i}$ 
in (\ref{actR}) or, equivalently, sets $\tilde{\partial}^{i}=0$, the action reduces to the 
conventional low-energy effective action 
 \be \label{10dimnsaction}
  S =  \int d^{D} x \sqrt{g} 
  e^{-2 \phi}  \left[ R + 4 (\partial \phi)^2  - \frac{1}{12} H^{ijk}H_{ijk}  \right] \; ,
 \ee
 where $H_{ijk}=3\partial_{[i}b_{jk]}$ is the field strength of the 2-form. 
 Moreover, for $\tilde{\partial}^{i}=0$ the gauge transformations (\ref{manifestH}) 
 with parameter $\xi^{M}=(\tilde{\xi}_{i},\xi^{i})$ reduce to the conventional 
 general coordinate transformations $x^{i}\rightarrow x^{i}-\xi^{i}(x)$ and 
 to the gauge transformations of the 2-form,  $\delta b_{ij}=2\partial_{[i}\tilde{\xi}_{j]}$. 

Let us now turn to the extension by the RR sector. 
In this we make significant use of the work of Fukuma, Oota, and 
Tanaka~\cite{Fukuma:1999jt}. (See also 
\cite{Hull:1994ys,Hassan:1999mm}.) 
The RR sector consists of forms 
of degrees 1 and 3 for type IIA and of degree 2 and 4 for type IIB, where the 5-form 
field strength of the 4-form is subject to a self-duality constraint. Here, we will 
use a democratic formulation that simultaneously uses dual forms, such that 
type IIA contains all odd forms, and type IIB contains all even forms, both 
being supplemented by duality relations \cite{Fukuma:1999jt}. 
The set of all forms naturally combines into a Majorana spinor of $O(10,10)$. 
Imposing an additional Weyl
condition yields a spinor containing either all even or all odd forms, 
and we will show that the DFT extension of the RR sector can be  
formulated in terms of such a spinor.

We start by fixing our conventions for the spinor representation,  
setting $D=10$ from now on. 
More precisely, these are representations of the double covering 
groups Pin$(10,10)$ of $O(10,10)$, and Spin$(10,10)$ of $SO(10,10)$. 
The gamma matrices satisfy the Clifford algebra 
 \be\label{Clifford}
  \{ \Gamma^{M},\Gamma^{N}\} = 2\eta^{MN}\,{\bf 1}\;.
 \ee
A convenient representation can be constructed using 
fermionic oscillators   
$\psi^i$ and $\psi_i$, satisfying
 \be \label{deffermosc}
  \{ \psi_i ,  \psi^j\} = \delta_{i}{}^{j}\;, \quad
  \{\psi_i,\psi_j\}  =   0\;, \quad \{\psi^i,\psi^j\} = 0\;,
 \ee
where $\left(\psi_i\right)^{\dagger}  = \psi^i$. With (\ref{ODDmetric})
we infer that they realize the algebra (\ref{Clifford}) via 
 \be \label{defgamma}
   \Gamma_{i}  =  \sqrt{2}\psi_{i}\;, \quad
   \Gamma^{i}  =  \sqrt{2}\psi^{i}\;.
 \ee
Introducing a `Clifford vacuum' $\ket{0}$ with $\psi_i\ket{0}=0$ for all $i$, 
and the normalization $\bra{0} 0 \rangle = 1$,
we can construct the representation by successive application of the 
raising operators $\psi^i$.  A general spinor state then reads 
 \be \label{genstate}
  \chi = \sum_{p=0}^{D}\frac{1}{p!}\,C_{i_1\ldots i_p}\,\psi^{i_1}\ldots\psi^{i_p}\ket{0}\;,
 \ee
whose coefficients $C_{i_1\ldots i_p}$ can be identified with 
$p$-forms $C^{(p)}$. 
Any element $S$ of the Pin group projects,  
via a group homomorphism $\rho:\;\,$Pin$(10,10)\rightarrow O(10,10)$,
to an element $h\in O(10,10)$, 
 \be\label{invgamma2}
  S\,\Gamma_{M}\,S^{-1}  =  \Gamma_{N}\, h^{N}{}_{M}\;, \quad
  h=\rho(S)\;,
 \ee
where $h\eta h^{T}=\eta$. Conversely, for any $h\in O(10,10)$, there 
is an $S\in {\rm Pin}(10,10)$ such that both  
 $\pm S$ project to~$h$. 
A spinor can be projected to a spinor of fixed 
chirality, i.e., to eigenstates  $\chi_{\pm}$ of $(-1)^{N_{F}}$
with eigenvalues $\pm 1$,  
where $N_{F}  =  \sum_{k}\psi^{k}\psi_{k}$ is the `fermion number operator'. 
The spinor $\chi_+$ of positive chirality then contains only even forms,
and the spinor $\chi_{-}$ of negative chirality contains only odd forms. 
Imposing a chirality constraint reduces the symmetry from Pin$(10,10)$ to Spin$(10,10)$
since only the latter leaves this constraint invariant. Finally, 
we need the charge conjugation matrix satisfying 
  \be\label{GammaunderC} 
   C \, \Gamma^M \, C^{-1} =
   (\Gamma^M)^\dagger  \;.
 \ee
A particular realization is given by 
 \be\label{Cmatrix}
  C= (\psi^{1}-\psi_{1})(\psi^{2}-\psi_{2})\cdots (\psi^{10}-\psi_{10})\;,
 \ee
which satisfies $ C\psi_{i}C^{-1} = \psi^{i}$ and thereby (\ref{GammaunderC}).

Given a spinor 
(\ref{genstate}) we can act with the Dirac operator
 \be\label{def-dir-op}
  \slashed\partial  \equiv   {1\over \sqrt{2}} \, \Gamma^M \partial_M
  =    \psi^i\partial_i+\psi_i\tilde{\partial}^{i} \,,
 \ee
which can be viewed as the $O(10,10)$ invariant extension of the 
exterior derivative $d$. In fact, for $\tilde{\partial}=0$, it differentiates 
with respect to $x^{i}$ and increases the form degree by one, 
thus acting like $d$. Moreover,  it squares to zero,
\be\label{zerosquare}
  \slashed{\partial}^2  =  
  \frac{1}{2}\Gamma^{M}\Gamma^{N}\partial_{M}\partial_{N} = 
  \frac{1}{2}\eta^{MN}\partial_{M}\partial_{N} =  0\,,
 \ee
using (\ref{Clifford}) and the constraint (\ref{ODDconstr}). 

In order to write an action that couples the NS-NS fields represented 
by the generalized metric ${\cal H}$  in (\ref{firstH}) to the RR fields represented 
by a spinor $\chi$, we note 
that the matrix ${\cal H}$ is an $SO(10,10)$ 
group element and thus has a representative in Spin$(10,10)$, 
as has been used in dimensionally reduced theories \cite{Fukuma:1999jt}.  
In our case, however, a subtlety arises because (\ref{firstH}) 
contains the full space-time metric, which we assume to be of Lorentzian signature. 
$SO(10,10)$ has two connected components, $SO^{+}(10,10)$, which contains 
the identity, and $SO^{-}(10,10)$.  
Due to the Lorentzian signature of $g$, ${\cal H}$ is actually an element of $SO^{-}(10,10)$. 
It turns out  
that a spin representative $S_{\cal H}\in\;$Spin$(10,10)$ of ${\cal H}$ 
cannot be constructed consistently over the space of all ${\cal H}$. 
For instance, one may find a closed loop ${\cal H}(t)$, $t\in[0,1]$, ${\cal H}(0)={\cal H}(1)$, 
in $SO^{-}(10,10)$, with the initial and final 
elements related by a \textit{time-like} T-duality, 
for which a continuously defined   
spin representative yields $S_{{\cal H}(1)}=-S_{{\cal H}(0)}$.
As a result, time-like T-dualities cannot be realized as transformations
of the conventional fields $g$ and $b$.  
Nevertheless, a  fully T-duality invariant action 
can  be written if we  treat the spin representative 
itself as the dynamical field. We thus introduce a field $\fancys$, satisfying
 \be\label{basicS}
   \fancys  =  \fancys^{\dagger}\;, \qquad \fancys  \in \text{Spin}^{-}(10,10)\;.
 \ee
The generalized metric is then \textit{defined} by the group homomorphism,  
$\rho(\fancys)={\cal H}$. 
By (\ref{basicS}) and the general properties of the group homomorphism \cite{ToAppear}, 
${\cal H}^{T}  =  \rho(\fancys^{\dagger})  =  {\cal H}$ and 
so, as required, ${\cal H}$ is symmetric. 

We are now ready to define the DFT formulation of type II theories, whose
independent fields are $\fancys$, $d$ and $\chi$. The action reads  
\be
\label{totaction}
 S  =   \int dx d\tilde x\, \Bigl(  e^{-2d}\, {\cal R} ({\cal H} , d)  + \frac{1}{4}
 (\slashed{\partial}{\chi})^\dagger \;
 \fancys \; \slashed\partial\chi\,\Bigr) \,,
\ee
and is supplemented by the self-duality constraint 
 \be\label{dualityintro}
 \slashed\partial\chi  = -{\cal K}\,\slashed\partial\chi\,, \quad 
 {\cal K}\equiv C^{-1} \fancys\, .
\ee
For the special case of type IIA, a similar duality relation has also been 
proposed in \cite{West:2010ev}.

The field equation of $\chi$ reads
  \be
   \slashed\partial \bigl( {\cal K}\,\slashed{\partial}\chi\bigr) =  0\;,
 \ee
which also follows as an integrability condition from the duality 
relation (\ref{dualityintro}), upon acting with $\slashed{\partial}$ and using  
(\ref{zerosquare}). The field equation of $\fancys$ reads
  \be
  {\cal R}_{MN}+{\cal E}_{MN}  =  0\;,
 \ee
where ${\cal R}_{MN}$ is the DFT extension of the Ricci 
tensor \cite{Hohm:2010jy},   
and the `energy-momentum' tensor reads, using (\ref{dualityintro}), 
  \be
 \label{rrstress-tensor+duality}
  {\cal E}^{MN}  =   - \tfrac{1}{16} {\cal H}^{(M}{}_{P}\,
  \overline{\slashed{\partial}\chi} \,
 \,\Gamma^{N)P}\,\slashed{\partial}\chi\;.
 \ee

Let us now discuss the symmetries of this theory. First, it is 
invariant under a global action by  $S\in{\rm Spin}^{+}(10,10)$, 
\be 
 \chi  \to  S \chi\;, \quad \fancys \to \fancys' =    (S^{-1})^\dagger\, \fancys\, S^{-1} \;,
\ee
implying $\slashed{\partial}\chi\rightarrow S\slashed{\partial}\chi$.
Specifically, $\chi$ is assumed to have a fixed chirality, which breaks 
the invariance group of the action from Pin$(10,10)$ to Spin$(10,10)$, 
while the duality relations break the invariance group to Spin$^{+}(10,10)$. 
The gauge symmetries of this theory are given by   
 \be
  \delta_{\lambda} \chi =  \slashed{\partial} \lambda\,,
 \ee
with spinorial parameter $\lambda$, leaving (\ref{totaction}) and (\ref{dualityintro})
manifestly invariant by (\ref{zerosquare}), and 
the gauge symmetry (\ref{manifestH}) parametrized by $\xi^{M}$. On 
the new fields $\fancys$ and $\chi$ it reads 
 \be\label{newxiM}
 \begin{split}
  \delta_\xi \chi &= \xi^M \partial_M  \chi +  {1\over 2}   \partial_M \xi_N \Gamma^M
  \Gamma^N \chi\;, \\
  \delta_{\xi} {\cal K}  &=  \xi^M \partial_M  {\cal K}
   + {1 \over 2} \big[ \Gamma^{PQ},   {\cal K} \, \big]  \partial_{P} \xi_{Q} \; ,
 \end{split} 
 \ee
where we have written the gauge variation of $\fancys$ in terms of ${\cal K}$ defined 
in (\ref{dualityintro}). It can be checked that 
this gauge transformation gives rise
to the required variation (\ref{manifestH}) of ${\cal H}$ upon application of $\rho$. 

We will now evaluate the DFT defined by (\ref{totaction})  and (\ref{dualityintro}) in 
particular T-duality `frames', starting with $\tilde{\partial}^{i}=0$. To this end, we
have to choose a particular parametrization of $\fancys$.
Writing 
  \be
   {\cal H} =   \begin{pmatrix} 1 &   0 \\ b & 1 \end{pmatrix}
   \begin{pmatrix} g^{-1} &   0 \\ 0 & g \end{pmatrix}
   \begin{pmatrix} 1 &   -b \\ 0 & 1 \end{pmatrix}  \equiv 
   h_b^T\,h_{g}^{-1}\,h_{b} \; ,
 \ee
we have to find spin representatives of the group elements $h_b$ and $h_{g}$. 
The subtlety here is that, with $g$ Lorentzian, $h_{g}$ takes values in 
$SO^{-}(10,10)$ and thus is not in the component connected to the identity. 
It is then convenient to write $g$ in terms of vielbeins, 
 \be
  g  =  e\,k \, e^{T}\;, \quad h_{g}=h_{e} h_{k} h_{e}^{T}\,, 
 \ee
where $e$ has positive determinant, i.e., $e\in GL^{+}(10)$, 
and $k$ is the flat Minkowski metric diag$(-1,1,\ldots,1)$.  
The group elements $h_{e}$ and $h_{b}$ are in the component 
connected to the identity and so their  
spin representatives can be written as 
simple exponentials, 
 \be \label{spin-elem}
    S_b  =     e^{- \frac{1}{2}b_{ij}\psi^{i}\psi^{j}}  \; , \quad 
   S_e   =    \frac{1}{\sqrt{\det e}} \, e^{ \psi^i E_i{}^j \psi_j } \, ,  
\ee
with $e = \exp (E)$, as can be verified with (\ref{invgamma2}). 
A spin representative for the matrix $k$ can be chosen to be \cite{Gualtieri}
 \be
  S_{k} = \psi^{1}\psi_{1}-\psi_{1}\psi^{1}\;, 
 \ee
where $1$ labels the time-like coordinate. This   
can also be verified with (\ref{invgamma2}). A spin representative
$S_{\cal H}$ of ${\cal H}$ can then locally be defined as 
\be
\label{definitionSH}
S_{\cal H}  \equiv   S_b^{\dagger}\,  S_{g}^{-1}\, S_{b} \;, \quad
S_{g} = S_{e}S_{k}S_{e}^{\dagger}\;.
\ee
We now set $\fancys=S_{\cal H}$, but we stress that this is just a particular parameterization 
in much the same way that (\ref{firstH}) is just a particular 
parametrization of ${\cal H}$. 

It is now straightforward to evaluate the action (\ref{totaction}) for $\tilde{\partial}=0$. 
First, as noted above, $\slashed{\partial}\chi$ reduces to the exterior derivatives of the $C^{(p)}$,
$F^{(p+1)}\equiv d C^{(p)}$.  The action of $S_{b}$ in $S_{\cal H}$ then modifies this, 
using (\ref{spin-elem}), to 
\be\label{ayvg}
  \widehat F =  e^{-b^{(2)}} \wedge F  = 
  e^{-b^{(2)}} \wedge dC  \,. 
\ee
Second, (\ref{definitionSH}) implies for the action of 
$S_{g}^{-1}$ 
\be
  S_{g}^{-1} \psi^{i_1}...\, \psi^{i_p}\ket{0}  =  -\sqrt{g} g^{i_1j_1}...\,
   g^{i_pj_p} \psi^{j_1}...\, \psi^{j_p} \ket{0}\;. 
\ee
The Lagrangian corresponding to the RR part of (\ref{totaction}) then reduces to 
kinetic terms for all forms,  
  \be
 \label{action-reduction}
  {\cal L}_{\rm RR} =  -\frac{1}{4}\sqrt{g}\,\sum_{p=1}^{D}\,\frac{1}{p!}\,g^{i_1 j_1}\cdots g^{i_p j_p}\,
  \widehat{F}_{i_1\ldots i_p}\widehat{F}_{j_1\ldots j_p}\;,
 \ee
where we recall that the sum extends over all even or all odd forms, depending on the 
chirality of $\chi$. Similarly, using (\ref{Cmatrix}), the self-duality constraint (\ref{dualityintro}) reduces 
to the conventional duality relations (with the Hodge star $*$), 
 \be\label{dualityRel}
  \widehat{F}^{(p)}  =  (-1)^{\frac{(D- p)(D- p-1)}{2}} *\widehat{F}^{(D-p)} \; .
 \ee
We have thus obtained the democratic formulation of type II theories, whose 
field equations are equivalent to the conventional field equations of type IIA for 
odd forms and of type IIB for even forms \cite{Fukuma:1999jt}. 

Let us briefly comment on the gauge symmetries for $\tilde{\partial}=0$.  
The transformations (\ref{newxiM}) for $\chi$, parameterized by $\xi^{M}=(\tilde{\xi}_{i},\xi^{i})$, 
reduce to the conventional general coordinate transformations $x^{i}\rightarrow x^{i}-\xi^{i}(x)$
of the $p$-forms $C^{(p)}$, but also to non-trivial transformations under the $b$-field gauge parameter 
$\tilde{\xi}_{i}$, $ \delta_{\tilde{\xi}} C =   d\tilde \xi \wedge C$.

We turn now to the discussion of other T-duality frames, starting with $\partial_{i}=0$, 
$\tilde{\partial}^{i}\neq 0$. For the analysis of 
this case it is convenient to perform a field redefinition according to the T-duality 
transformation $J$ that exchanges $x^{i}$ and $\tilde{x}_{i}$ and which, as a matrix, 
coincides with $\eta$ defined in (\ref{ODDmetric}), 
  \be\label{Hdual}
  {\cal H}^{\prime}  \equiv 
  J\,{\cal H}\,J  =  {\cal H}^{-1} \;.
 \ee
It has been shown in \cite{Hohm:2010jy} that the NS-NS part of the DFT reduces 
for $\partial_{i}=0$ to the same action (\ref{10dimnsaction}), but written in terms of
the primed (T-dual) variables. Next, we define a corresponding field redefinition 
for the RR fields, using a spin representative $S_{J}$ 
of $J$, 
 \be\label{redefD}
  \chi^{\prime} =  S_{J}\chi\,, \quad \slashed{\partial}^{\prime}
  =\psi^{i}\tilde{\partial}^{i}+\psi_i\partial_i \,, \quad
  \slashed{\partial}^{\prime}\hspace{-0.1ex}\chi^{\prime}=S_{J}\slashed{\partial}\chi\,.
 \ee
For the RR action we then find   
\be\label{redefaction}
 \begin{split}
    {\cal L}_{\rm RR} &= \tfrac{1}{4}(\slashed{\partial}\chi)^{\dagger}S_{\cal H} \slashed{\partial}\chi
     =  \tfrac{1}{4}(\slashed{\partial}^{\prime}\hspace{-0.1ex}
     \chi^{\prime})^{\dagger}(S_{J}^{-1})^{\dagger}S_{\cal H}S_{J}^{-1}
    \slashed{\partial}^{\prime}\hspace{-0.1ex}\chi^{\prime} \\
     &=  -\tfrac{1}{4} (\slashed{\partial}^{\prime}\hspace{-0.1ex}\chi^{\prime})^{\dagger}S_{{\cal H}^{\prime}}
    \slashed{\partial}^{\prime}\hspace{-0.1ex}\chi^{\prime} \;,
  \end{split}
 \ee
where we used that $J$ contains a time-like T-duality such that, as mentioned above, this 
leads to a sign factor in the transformation of $S_{\cal H}$. Thus, in the 
new variables the action takes the same form as in the original variables, up to a sign. 
The transformed Dirac operator in (\ref{redefD}) implies that 
setting $\partial_{i}=0$ in the first form in (\ref{redefaction}) is equivalent 
to setting $\slashed{\partial}^{\prime}=\psi^{i}\tilde{\partial}^{i}$ in the final form in  (\ref{redefaction}). 
This way to evaluate the action is, however, equivalent to our computation above 
of setting $\tilde{\partial}=0$ in the original action, just with fields and derivatives replaced 
by primed fields and derivatives. Thus, we conclude that the DFT action reduces for 
$\partial_{i}=0$ to a type II theory with the overall sign of the RR action reversed.  
These are known as type II$^{\star}$ theories and have been introduced by 
Hull in the context of \textit{time-like} T-duality  \cite{Hull:1998vg}. They are 
defined such that the time-like circle reductions of type IIA (IIB) and type 
IIB$^{\star}$ (IIA$^{\star})$ are equivalent.  
This result also implies that the overall sign of $\fancys$ has no physical significance 
in that it merely determines for which coordinates ($x$ or $\tilde{x}$) we obtain the 
type II or type II$^{\star}$ theory. 

More generally, one finds that evaluating the DFT in a T-duality frame that is obtained 
by an odd (even) number of T-duality inversions 
from a frame in which the theory reduces, say, to type IIA, it reduces to the 
T-dual theory, i.e., type IIB (IIA) for space-like transformations and IIB$^{\star}$ (IIA$^{\star})$ 
for time-like transformations.   
Summarizing, the DFT defined by (\ref{totaction})  and (\ref{dualityintro}) 
combines all type II theories in a single universal formulation. 
We hope that this theory may provide insights into the still elusive 
formulation of string theory as, e.g., for a yet to be constructed type II string field 
theory. \vspace{0.2cm}

\begin{acknowledgments}\vspace{-0.45cm}
We thank  M.~Gualtieri, J.~Maldacena, A.~Sen and P.~Townsend for useful
conversations.  We thank C. Hull for comments,
and understand from him that he has worked in
closely related directions.
This work is supported by the 
DoE (DE-FG02-05ER41360). OH is supported by the DFG -- The German
Science Foundation, and SK is supported in part by a Samsung Scholarship. 
\end{acknowledgments}

\end{document}